# Emergence of the Macroscopic Fourier Law from the Microscopic Wave Equation of Diffusive Medium

Er'el Granot, Nisim Cohen and Shmuel Sternklar,

*Department of Electrical and Electronics Engineering, Ariel University Center of Samaria, Ariel 44837, Israel*

## Abstract

The Fourier law and the diffusion equation are derived from the Schrödinger equation of a diffusive medium (consisting of a random potential). The theoretical model is backed by numerical simulation. This derivation can easily be generalized to demonstrate the transition from any random wave equation to the diffusive equation.

## Introduction

The well-known Fourier law relates the heat current $J$ to the gradient of the temperature $\tilde{T}$. In an effectively 1D systems, this law can be written

$$J = -\kappa \frac{\partial}{\partial x} \tilde{T} . \qquad (1)$$

where $\kappa$ is the material heat conductivity. Despite of the fact that this relation is intuitively appealing, and in most textbooks it appears without an appropriate explanation, its derivation from microscopic considerations has always been a challenge. In fact, ever since its first appearance its validity was proven in many experiments. However, we are not aware of satisfactory theoretical proofs or arguments, which establish its' emergence from basic microscopic behavior. There have been some attempts to understand such a transition (see, for example, Refs.1-5), however, a full understanding of the transition is still lacking and requires additional work.

The Fourier law is a particular example of a much broader group of diffusive processes. This law can be applied to any diffusion mechanism. It relates the (particles) density $\rho$ gradient to the current, while the proportionality constant is the diffusion coefficient $D$

$$J = -D\nabla\rho \tag{2}$$

In general, the diffusion coefficient can be written as

$$D = \frac{v}{\mu_s} \tag{3}$$

where $\mu_s$ is the scattering coefficient and $v$ is the propagation velocity in the medium.

In the 1D case, the diffusion coefficient (in its conductivity form it is related to Landauer equation[6-8]) is

$$D = av\frac{T}{1-T} \tag{4}$$

where $a$ is the mean distance between adjacent scatterers, $T$ is their mean transmission coefficient and as above $v$ is the mean velocity of the particles. Therefore, in our case $\mu_s = (1-T)/aT$.

One of the remarkable things about the Fourier law is that presumably it contradicts its' microscopic origin. The microscopic dynamics of many random processes is governed by some sort of wave equation (e.g. the Maxwell wave equation, the acoustic wave equation, the Scrödinger equation etc.); however, if the Fourier law is integrated in the continuity equation

$$\frac{\partial \rho}{\partial t} + \nabla J = 0$$

the diffusion equation emerges

$$\frac{\partial \rho}{\partial t} - D\nabla^2 \rho = 0.$$

The difference between the diffusion equation and the original wave equation, which describes the microscopic dynamics, is fundamental. The information on the medium, which appears in the wave equation as an additive but very complex term, somehow converges into a single coefficient $D$ of the Laplacian term. The temporal derivatives $\partial^2/\partial t^2$ or $i\partial/\partial t$ (in the Schrödinger case) totally disappear and a new $\partial/\partial t$ term emerges. Therefore, the solutions of the microscopic dynamics are totally different from the solutions of the diffusion equation.

In the conductivity case, the heat, which is a form of energy, is transferred through the medium by the particles' mobility. The randomness of the process is caused by random positions and directions of the bouncing particles. If, however, the quantum particles are transferred from one end of a diffusive medium to the other, the random locations of the scatterers are the source of the randomness in the process. This is what makes the medium diffusive. The quantum particles can be either electrons, photons, phonons or any other quantum particles. This scenario is much simpler to analyze, since it can be solved by a stationary-state wave equation (either the Schrödinger or the Maxwell's wave equation). This is the route we choose to take in this paper.

The problem of diffusion in one-dimensional systems raised a lot of attention since Anderson's well-known work on localization[9] (see for example Refs. 10-12). However, to the best of our knowledge, these works did not include the transition from the microscopic to the macroscopic regimes, and did not include calculations of the diffusion coefficient from microscopic parameters. We will demonstrate that the macroscopic diffusion equation (and the Fourier law) can be derived from the microscopic wave (or the Schrödinger) equation. We will focus on the dynamics of a quantum particle in a diffusive medium (potential), however, it can easily be generalized to any wave equation.

We begin with the Schrödinger equation of a diffusive medium

$$-\frac{\partial^2}{\partial x^2}\psi(x,t)+V(x,t)\psi(x,t)=i\dot\psi(x,t) \tag{5}$$

where we used the units $\hbar=1$ and $2m=1$ ($m$ is the particles' mass and $\hbar$ is the Planck constant divided by $2\pi$), and $V(x,t)$ is the medium's random potential, which varies in time.

We assume that the potential varies so slowly in time that the dynamics can be represented by a stationary-state wave equation. That is, we freeze the time at specific instants $t_j$ and calculate the wavefunction for a static potential

$$-\frac{\partial^2}{\partial x^2}\psi(x,t_j)+[V(x,t_j)-\omega]\psi(x,t_j)=0 \tag{6}$$

where $\omega$ is the particles' energy (or angular frequency).

After repeating this process for many instances $t_j$, we average over the different $|\psi(x,t_j)|^2$. This process is equivalent to averaging a slow process over extremely long times.

## Theory

First we show that after adiabatic averaging over the locations of the scatterers the phases of the wavefunction can be ignored, and in fact the dynamics is totally governed by the *amplitude* of the wavefunctions as should be expected from a dephasing process. As a result, the problem reduces to a semi-classical one.

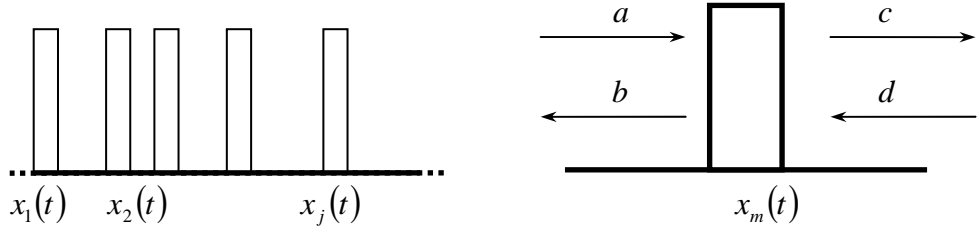

Fig.1: a) An illustration of the diffusive medium with multiple scatterers: b) the scattering coefficients of a single scatterer.

We assume that the random potential consists of multiple scatterers, each of which can be characterized by a certain potential barrier (or well), as shown in Fig.1a. For simplicity we assume that the barriers are identical. For a given wavenumber $k$ the wavefunction at the left side of a scatterer (Fig. 1b) is

$$\psi_{left} = a\exp(ikx) + b\exp(-ikx) \qquad (7)$$

and on its right side,

$$\psi_{right} = c\exp(ikx) + d\exp(-ikx) \qquad (8)$$

If $\tilde{\tau}$ and $\tilde{\rho}$ are the transmission and reflection coefficients of a single barrier (scatterer) respectively, then

$$c = a\tilde{\tau} + d\tilde{\rho}\exp(-i\varphi) \qquad (9)$$
$$b = a\tilde{\rho}\exp(i\varphi) + d\tilde{\tau} \qquad (10)$$

where $\varphi = 2kx_m$ is an additional phase that depends on the exact location of the scatterer.

Therefore, the absolute values of the coefficients are

$$|c|^2 = |a|^2 T + |d|^2 R + a\tilde{\tau}d^*\tilde{\rho}^*\exp(i\varphi) + \text{c.c.} \qquad (11)$$

$$|b|^2 = |a|^2 R + |d|^2 T + a\tilde{\rho}d^*\tilde{\tau}^* \exp(i\varphi) + \text{c.c.} \tag{12}$$

where $T \equiv |\tilde{\tau}|^2$ and $R \equiv |\tilde{\rho}|^2$ and c.c. stands for complex conjugate.

Now, by averaging over an ensemble of random media, where the position of the barriers varies, the coefficients $a, b, c$ and $d$ vary as well, and so does the phase $\varphi$. The particles' density

However, due to the complexity of the medium and its multiple degrees of freedom, there is no correlation between the phase and the coefficients. Therefore, the two last terms in both equations vanish after such ensemble averaging, yielding

$$\langle |c|^2 \rangle = \langle |a|^2 \rangle T + \langle |d|^2 \rangle R \tag{13}$$

$$\langle |b|^2 \rangle = \langle |a|^2 \rangle R + \langle |d|^2 \rangle T \tag{14}$$

Therefore, we can connect the *average* value of the coefficients *amplitude* at the right side of the barrier to their value at its left side by a simple matrix

$$\begin{bmatrix} \langle |c|^2 \rangle \\ \langle |d|^2 \rangle \end{bmatrix} = \frac{1}{T} \begin{bmatrix} T-R & R \\ -R & 1 \end{bmatrix} \begin{bmatrix} \langle |a|^2 \rangle \\ \langle |b|^2 \rangle \end{bmatrix} \tag{15}$$

Since the phase is lost after averaging the exact location of each scatterer is insignificant, and only the number of scatterers matters. Thus, after $n$ scatterers the entire medium can be characterizes by the matrix (which relates the right-most coefficients to the left-most ones)

$$\frac{1}{T^n}\begin{bmatrix} T-R & R \\ -R & 1 \end{bmatrix}^n = \frac{1}{T}\begin{bmatrix} T-nR & nR \\ -nR & T+nR \end{bmatrix} \tag{16}$$

If, at a given point the average particle density is

$$\rho(x) \equiv \langle |\psi(x)|^2 \rangle = \langle |a|(x)^2 \rangle + \langle |b|(x)^2 \rangle \qquad (17)$$

[where here $a(x)$ and $b(x)$ are coefficients of the local $\exp(ikx)$ and $\exp(-ikx)$ terms respectively at $x$], and the local current is

$$J = 2k \left( \langle |a|^2 \rangle - \langle |b|^2 \rangle \right) \qquad (18)$$

(where again the units $\hbar = 2m = 1$ where used), then, after a distance $L$, which is equivalent to passing through $n = L/a$ scatterers, where $a$ is the mean distance between adjacent scatterers, the average coefficients amplitudes satisfy

$$\begin{bmatrix} \langle |a(L)|^2 \rangle \\ \langle |b(L)|^2 \rangle \end{bmatrix} = \frac{1}{T} \begin{bmatrix} T - nR & nR \\ -nR & T + nR \end{bmatrix} \begin{bmatrix} \langle |a(0)|^2 \rangle \\ \langle |b(0)|^2 \rangle \end{bmatrix} . \qquad (19)$$

From Eq. 19 two immediate consequences emerge: the current density is, of course, a constant throughout the medium $j(L) = j(0)$, but the density varies according to

$$\rho(L) = \langle |a(L)|^2 \rangle + \langle |b(L)|^2 \rangle = \left[ \left( 1 - 2n \frac{R}{T} \right) \langle |a(0)|^2 \rangle + \left( 1 + 2n \frac{R}{T} \right) \langle |b(0)|^2 \rangle \right] \qquad (20)$$

and therefore,

$$\rho(L) - \rho(0) = -2n \frac{R}{T} \frac{J}{2k} \qquad (21)$$

by substitution $n = L/a$, we obtain

$$\frac{\rho(L) - \rho(0)}{L} = -\frac{R}{T} \frac{1}{ak} J, \qquad (22)$$

which shows the linear relation between the density gradient and the current, i.e.,

$$D \frac{\partial \rho}{\partial x} = J \qquad (23)$$

where the dispersion coefficient in this case is

$$D = ak\frac{T}{R}. \tag{24}$$

Reverting to ordinary physical units, it becomes $D = \frac{\hbar}{2m} ak \frac{T}{R}$.

Moreover, since Eq.21 is valid for any $L$ the density varies linearly thought the medium

$$\rho(x) = \rho(0) - \frac{R}{T}\frac{1}{ak} Jx. \tag{25}$$

These results clearly show the emergence of the 1D Fourier law (eq. 23), and is definitely a solution of the stationary-state 1D diffusion equation.

## Simulation

To simulate the varying random medium we assume, for simplicity, that it consists of $M$ randomly distributed identical delta-function potentials, whose positions changes adiabatically. For any practical purpose a delta-function potential is an excellent approximation for any scatterer whose width is considerably smaller than the incoming particles' wavelength. Moreover, this derivation and simulation can easily be applied to any scatterer as eqs.7-25 suggest.

Therefore, the potential of Eq.6 can be written

$$V(x,t_j) = \sum_{m=1}^{M} \alpha\delta[x - x_m(t_j)] \tag{26}$$

In this case, the transmission coefficient of a single scatterer is [13] $T = \frac{1}{1+(\alpha/2k)^2}$ and therefore (according to Eq.24) the theoretical dispersion coefficient is

$$D = 4a\frac{k^3}{\alpha^2}. \tag{27}$$

Again, with ordinary physical units, this should read $D = 2a \dfrac{\hbar}{m} \dfrac{k^3}{\alpha^2}$.

In Figs.2 and 3 the particle density $\rho(x)$ is plotted vs. the longitudinal coordinate for different sample averaging. In this simulation we took $\alpha = 1$, $k = 2\pi$ $N = 100$ and $a$ was chosen a uniformly distributed random variable with the average value $\langle a \rangle = 0.75$.

As can be seen from Fig. 2, in the case of a single medium (i.e. no averaging) the density fluctuates considerably through the medium and it seems like a totally random function. However, after averaging over $10^5$ random samples, the linear dependence on the distance is clearly seen (see Fig.3).

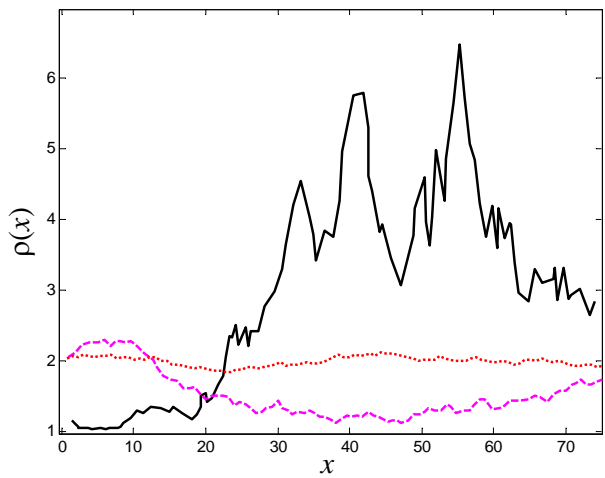

Fig.2: The distribution of particle density $\rho(x)$ in the medium for a single medium (black solid line), averaging over 10 random samples (magenta dashed line) and averaging over 100 samples (red dotted line).

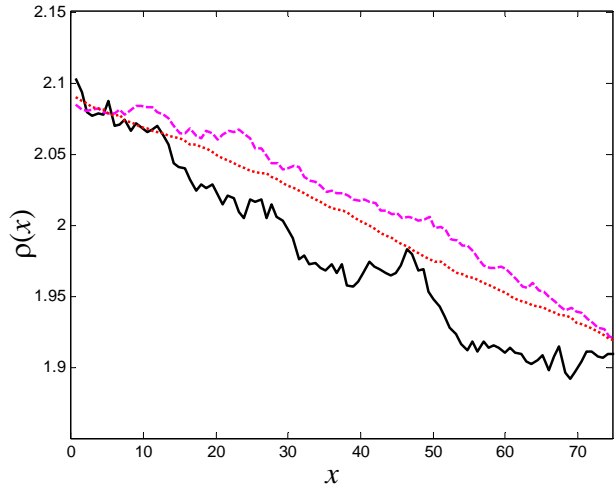

Fig.3: The distribution of particle density $\rho(x)$ in the medium while averaging over $10^3$ random samples (black solid line), averaging over $10^4$ samples (magenta dashed line) and averaging over $10^5$ samples (red dotted line).

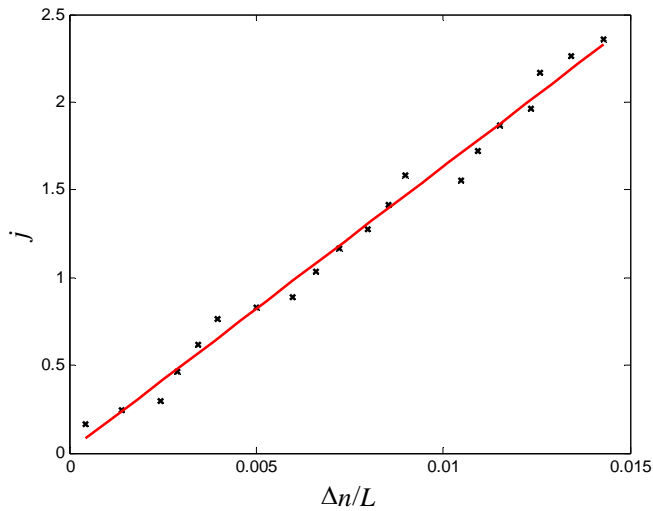

Fig.4: The current density $j$ vs. the particle gradient $\Delta n / L$ after averaging over $3 \times 10^4$ samples, each of which consist of $M = 40$ scatterers, with a scatterer strength $\alpha = 2$ and a mean distance between scatterers of $0.75$.

In Fig.4 we calculate the current density $j$ for different density differences $\Delta n \equiv \rho(0) - \rho(L)$. In this simulation there were $M = 40$ scatterers, their strength was $\alpha = 2$, the mean distance between adjacent scatterers was $\langle a \rangle = 0.75$ and the averaging was over $3\times 10^4$ random samples. The simulation prediction was $D_s = 162 \pm 30$, which agrees with the theoretical prediction $D_t \cong 186$.

## Summary

We presented a simple model of a diffusive medium, which is governed by the Schrödinger dynamics. This model was used to derive the macroscopic Fourier law, and therefore the diffusion equation, from ensemble averaging over microscopic quantum samples. That is, despite the fact that the particles' distribution within the diffusive medium was calculated exactly by the microscopic quantum equation, it was proven that the diffusion dynamics can be retrieved by ensemble averaging. The diffusion coefficient was derived theoretically, and was backed up by numerical simulation.

This research was partially supported by The Israel Science Foundation (grant No.144/03-11.6)